# Comparative study of the high-pressure behavior of $ZnV_2O_6$, $Zn_2V_2O_7$, and $Zn_3V_2O_8$

D. Díaz-Anichtchenko[1], D. Santamaria-Perez[1], T. Marqueño[1], J. Pellicer-Porres[1],

J. Ruiz-Fuertes[2], R. Ribes[1], J. Ibañez[3], S. N. Achary[4], C. Popescu[5], and D. Errandonea[1]

[1]Departamento de Física Aplicada-ICMUV, Universidad de Valencia, Dr. Moliner 50, Burjassot, 46100 Valencia, Spain

[2]DCITIMAC, MALTA Consolider Team, Universidad de Cantabria, 39005 Santander, Spain

[3]Institute of Earth Sciences Jaume Almera, CSIC, 08028 Barcelona, Spain

[4]Chemistry Division, Bhabha Atomic Research Centre, Trombay, Mumbai 400085, India

[5]CELLS-ALBA Synchrotron Light Facility, Cerdanyola, 08290 Barcelona, Spain

**Abstract:** We report a study of the high-pressure structural behavior of $ZnV_2O_6$, $Zn_2V_2O_7$, and $Zn_3V_2O_8$, which has been explore by means of synchrotron powder x-ray diffraction. We found that $ZnV_2O_6$ and $Zn_3V_2O_8$ remain in the ambient-pressure structure up to 15 GPa. In contrast, in the same pressure range, $Zn_2V_2O_7$ undergoes phase transitions at 0.66, 2.97, and 10.75 GPa. Possible crystal structures for the first and second high-pressure phases are proposed. Reasons for the distinctive behavior of $Zn_2V_2O_7$ are discussed. The compressibility of the different polymorphs has been determined. The respond to pressure is found to be anisotropic and room-temperature equations of state have been determined. The bulk modulus of $ZnV_2O_6$ (129 GPa) and $Zn_3V_2O_8$ (120 GPa) are consistent with a structural framework composed of $ZnO_6$ compressible octahedral and $VO_4$ uncompressible tetrahedral. In contrast $Zn_2V_2O_7$ is highly compressible with a bulk modulus of 58 GPa. This and the sequence of structural transitions are related to the fact than Zn atoms are coordinated to five oxygen atoms in this compound. A comparison to the high-pressure behavior of related compounds is presented.

**Introduction**

Vanadium oxide-based materials are receiving substantial attention due to their exceptional physical and chemical characteristic which make them useful for many technological applications. In particular, zinc metavanadate ($ZnV_2O_6$), zinc pyrovanadate ($Zn_2V_2O_7$), and zinc divanadate ($Zn_3V_2O_8$) have recently received a great deal of attention due to their proposed use in photocatalytic water splitting [1, 2]. The three photocatalytic materials possess favorable band-gap energies and appropriate band-edge positions for inducing the dissociation of water into its constituent parts, $H_2$ and oxygen $O_2$, using solar radiation. These materials also have applications as electrodes in lithium-ion batteries [3] and supercapacitors [4], as phosphor for white light-emitting diodes [5], and as long-life cathode materials of aqueous zinc-ion batteries [6].

In addition of the technological interest on different zinc vanadates, there is also a fundamental interest of them. Their diverse stoichiometry gives the possibility to make a systematic compassion of their crystal structures contributing to deepen the understanding of the crystal chemistry of vanadates and related compounds. $ZnV_2O_6$ crystallizes in a monoclinic crystal structure (space group C2/m) [7], $Zn_2V_2O_7$ is also monoclinic (space group C2/c- α-phase) [8], and the structure of $Zn_3V_2O_8$ is orthorhombic (Cmca) [8]. The structure mainly differs in the way that Zn and V coordination polyhedra are linked. Regarding the stabilities of these structures under extremal stimuli such as temperature and pressure, the existence of a high-temperature polymorph has been only reported for $Zn_2V_2O_7$ having it a monoclinic structure (space group C2/m- β-phase) [9]. Under high-pressure conditions, only $ZnV_2O_6$ has been studied. Density-functional theory calculations have predicted the occurrence of a phase transition towards an orthorhombic structure at 5 GPa [10]. However, x-ray diffraction (XRD) studies did not find evidence of this transition, being the ambient-pressure polymorph stable up to 16 GPa [11]. Motivated by this contradiction, and by the broad interest on zinc vanadates, we have performed a comparative study of the high-pressure behavior of $ZnV_2O_6$, $Zn_2V_2O_7$, and $Zn_3V_2O_8$. In particular, their structural stability has been studied by synchrotron powder XRD. We found that ambient-pressure polymorph of $ZnV_2O_6$ and $Zn_3V_2O_8$ remain stable up to 15 GPa. In contrast, $Zn_2V_2O_7$ undergoes at least three phase transitions in the same pressure. A detail comparison of the crystal structures and compressibilities of different zinc vanadates will be reported. Results will be also compared with the literature. Reasons for the distinctive behavior of $Zn_2V_2O_7$ will be proposed. Finally, candidate crystal structures of the high-pressure polymorphs of $Zn_2V_2O_7$ will be presented.

**Experimental details**

$ZnV_2O_6$ was synthesized from a 50–50 mol. % mixture of high-purity powders of ZnO and $V_2O_5$ following the method described by Quiñones-Galvan *et al.* [12]. $Zn_2V_2O_7$ was synthesized by the solid-state reaction method. $5ZnO \cdot 2CO_3 \cdot 4H_2O$ and $V_2O_5$ were mixed with stoichiometric molar ratio by thorough grinding in an agate mortar and heated in at furnace at 600 ºC for 6 h. $Zn_3V_2O_8$ was synthesized using the hydrothermal method. The starting materials were $Zn(CH_3COO)_2 \cdot 2H_2O$ and $V_2O_5$ mixed with a 1:1 molar ratio and heated in an autoclave at 250 ºC for 12 hours. All the samples were characterized by powder XRD using Cu K$\alpha$ radiation. Single phase samples with the crystal structure reported in the literature [7, 8] were confirmed for them.

High-pressure powder angle-dispersive XRD measurements were carried out at room-temperature using the MSPD-BL04 beamline at the ALBA synchrotron [13]. A wavelength of 0.4246 Å (0.4642 Å) and a beam size of 15 μm × 15 μm at ALBA, was used for $ZnV_2O_6$ and $Zn_3V_2O_8$ ($Zn_2V_2O_7$). XRD patterns were collected on a Rayonix charge-coupled device. A rocking (±3°) of the diamond-anvil cell (DAC) was used to reduce preferred orientations effects. The experiments were performed using DACs equipped with diamonds with 500-μm culets. The sample was loaded in a 200-μm diameter chamber drilled in stainless-steel gaskets, pre-indented to a thickness of 50 μm. All experiments were carried out using 16:3:1 methanol-ethanol-water as pressure medium. In $Zn_3V_2O_8$ a second experiment was performed using neon as pressure medium. Special care was taken during sample loading to prevent sample bridging between diamonds [14]. Pressure has been determined, with an error smaller than 0.05 GPa, using the ruby scale. In $Zn_2V_2O_7$ we also used the equation of state of copper [15]. The pressure determined by the two methods agree within the error.

**Crystal structure comparative**

This Section aims to describe the crystal structures of the four $Zn^{II}_XV^V_2O_{5+X}$ vanadates reported in the literature. In this work, we have studied the high-pressure structural behavior of three of them ($ZnV_2O_6$, $Zn_2V_2O_7$, and $Zn_3V_2O_8$) and the initial ambient conditions structures perfectly matched those previously published [2,10]. For the sake of completeness, we also include the description of a fourth existing vanadate, $Zn_4V_2O_9$.

**$ZnV_2O_6$**

The room-conditions crystal structure of $ZnV_2O_6$ was firstly determined by Angenault and Rimsky in 1968 and described within the C2 monoclinic space group [1]. Later, Andreetti et al. refined this structure using the centrosymmetric C2/m space group [7], like in other $MV_2O_6$ (M = Mg, Ca, Mn, Co and Cd) compounds [17-20]. The lattice parameters derived from our powder XRD measurements are $a$ = 9.265(1) Å, $b$ = 3.5242(5) Å, $c$ = 6.5889(7) Å, and $\beta$ = 111.37(2)°, in good agreement with previous experimental and theoretical data [1,7,10]. Our powder patterns have some spotty appearance due to texture effects, but the intensities of the diffraction maxima are qualitatively similar to those observed in previous experiments [7]. This fact suggests that the atomic coordinates reported in Ref. [2] are also those of our sample. The crystal structure consists of edge-shared [ZnO$_6$] octahedral chains (Zn – O distances: 2 apical x 1.981 Å + 4 equatorial x 2.244 Å) and distorted corner-shared [VO$_4$] tetrahedral chains (V – O distances: 1.659 Å + 1.681 Å + 2 x 1.856 Å), both running parallel to the *b*-axis. The V tetrahedral geometry entails that a fifth O atom could accommodate at a distance of 2.078 Å, and the existence of a sixth O neighbor at 2.573 Å yields highly distorted [VO$_6$] octahedra that would form edge-sharing layers perpendicular to the *c* direction. Figure 1 shows both the [ZnO$_6$] octahedra and [VO$_4$] tetrahedra connectivity. An explanation for the existence of these unbranched single chains of two-connected [VO$_4$] tetrahedra is provided by the extended Zintl-Klemm concept [21-24]. According to this qualitative approach, the Zn atoms would donate their two valence electrons to the vanadate anion [V$_2$O$_6$]$^{2-}$ converting it into pseudo (ψ-) [Cr$_2$O$_6$]. Consequently, the topology of the [V$_2$O$_6$]$^{2-}$ anion is the same as that of the CrO$_3$ chromium oxide [24].

**Zn$_2$V$_2$O$_7$**

The ambient-conditions structure of Zn$_2$V$_2$O$_7$ (α-phase) was solved and refined by Gopal and Calvo in 1972 [8]. These authors reported a C2/c monoclinic structure with lattice parameters *a* = 7.429(5) Å, *b* = 8.340(3) Å, *c* = 10.098(3) Å, and β = 111.37(5)°, which are in good agreement with those obtained by us from a Rietveld refinement of powder XRD data (*a* = 7.428(5) Å, *b* = 8.329(6) Å, *c* = 10.089(3) Å, and β = 111.29(9)°). In this structure the Zn atoms are coordinated to five O atom neighbors (penta-coordinated) and forms chains of distorted trigonal bipyramids that share edges perpendicularly along the [110] direction. The five Zn – O distances of these polyhedral units are between 1.97 and 2.09 Å, the next closest O atom lying at 3.35 Å. The V atoms are tetrahedrally coordinated and each pair of [VO$_4$] tetrahedra is linked by a common O atom, a common corner, to form the [V$_2$O$_7$]$^{4-}$ pyrovanadate anions aligned with the *a*-axis. Such type of [X$_2$O$_7$]$^{n-}$ anions are present in a large number of minerals [22,25,26]. The extended Zintl-Klemm concept suggests that the two zinc atoms per formula unit would donate four electrons to the [V$_2$O$_7$] group, which would be converted into [V$_2$O$_7$]$^{4-}$ ($\Psi$-Mn$_2$O$_7$) and adopt the same conformation as Mn$_2$O$_7$ itself [27].

**Zn$_3$V$_2$O$_8$**

The ambient-conditions form of this compound was refined by Gopal and Calvo from single crystal XRD data [8] and was described with an orthorhombic space group Cmca. Our room-pressure experimental results agree well with the lattice parameters and atomic coordinates previously reported: $a = 6.037(1)$ Å, $b = 11.438(2)$ Å and $c = 8.221(1)$ Å. The structure is based upon a close packing arrangement of O atoms, where the Zn atoms occupy octahedral sites and the V atoms occupy tetrahedral positions. Each [ZnO$_6$] octahedron shares edges with three adjacent octahedra and form corrugated layers perpendicularly to the [010] direction. These layers are interconnected by isolated orthovanadate [VO$_4$] units. In terms of the aforementioned extended Zintl-Klemm concept, the isolation of these units can be understood if we assume that the Zn atoms transfer of three electrons to the V atoms. This hypothetical V$^{3-}$ anion would have four pairs of electrons in its valence shell that could be approached by four O atoms to form an [VO$_4$]$^{3-}$ anion with closed electronic configuration.

**Zn$_4$V$_2$O$_9$**

The crystal structure of the metastable Zn$_4$V$_2$O$_9$ vanadate was reported in 1986 by Walburg and Müller-Buschbaum [28]. At ambient conditions, it was described with a monoclinic P2$_1$ space group and lattice parameters: $a = 10.488(5)$ Å, $b = 8.198(6)$ Å, $c =$ 9.682(5) Å, and β = 118.66(4)°. This unit-cell contains 16 Zn atoms, all of them penta-coordinated except two that are tetrahedrally coordinated. The [$ZnO_5$] triangular bipyramids share edges and corners, forming layers that lie parallel to (001). The V atoms are also tetrahedrally coordinated and, together with [$ZnO_4$] tetrahedra, they occupy the space between these layers. Walburg and Müller-Buschbaum merely described the structure in terms of isolated [$VO_4$] tetrahedra. Nevertheless, the crystal structure of $Zn_4V_2O_9$ can be elegantly reinterpreted on the basis of the extended Zintl-Klemm concept, providing important structural information. An important characteristic of this compound is the above-mentioned existence of two kinds of coordination polyhedra for Zn atoms. This fact suggests that zinc has an amphoteric character so that some Zn atoms (14/16) act as donors of electrons and hence are penta-coordinated, whereas others (2/16) behave formally as acceptors, adopting a tetrahedral coordination by O atoms. According to this interpretation, the chemical formula of this vanadate can be rewritten as $Zn_{14}[ZnV_2O_{10}]_2[VO_4]_4$. Therefore, the structure of this compound consists of a mixture 2:1 of orthovanadate anions [$VO_4$]$^{3-}$ and groups of three corner-connected tetrahedra forming discrete [$ZnV_2O_{10}$]$^{8-}$ anions. This topology can be understood in the light of the Zintl-Klemm concept if we consider that the penta-coordinated Zn atoms donate 14 x 2 = 28 electrons to both the Zn and V tetra-coordinated atoms. Twelve of these electrons would go to the four orthovanadate [$VO_4$]$^{3-}$ anions. The remaining 16 electrons would be transferred to the Zn and V atoms of the 2 [$ZnV_2O_{10}$]$^{8-}$ anions. Each central Zn atom would receive 4 electrons, having then 6 valence electrons (like S) and the two terminal V atoms of these discrete anions would receive 2 electrons each, converting into atoms with 7 valence electrons (like Cl). The skeleton of this anion consequently adopts a geometry similar to that of the real $SCl_2$ molecule. The O atoms are located close to the midpoint of the V – Zn contacts and the alleged lone electron pairs. It is worth mentioning that the coexistence of isolated tetrahedral units and discrete unbranched groups of 3 tetrahedra has been observed in several silicate minerals [29-31].

**Results and Discussion**

**ZnV$_2$O$_6$**

In Fig. 5 we present a selection of XRD patterns measured in ZnV$_2$O$_6$ at different pressures. The pattern measured at ambient pressure can be undoubtedly assigned to the known crystal structure which has been described in the previous section. This is illustrated by the small residuals of the Rietveld refinement shown in Fig. 5. As pressure increase, the peaks shift towards higher angles, as expected. However, different peaks evolve in a different way, as can be seen in the triplet present around 8º at ambient pressure. In spite of it, we found that all measured XRD patterns can be assigned to the ambient pressure structure. In Fig. 5, it can be seen that the pattern measured at 14.5 GPa (the highest pressure reached in the experiments) can be refined with small residuals assuming the structure of the ambient pressure polymorph and using only the unit-cell parameters as free parameters. This confirms the conclusions extracted by Tang et al. from Raman and XRD experiments [11]. In particular, no evidence of the theoretically predicted phase transition (at 5 GPa) [10] is found in our experiments. It should be interesting to explore in the future is the theoretically predicted pressure-induced transition is not observed due to the presence of kinetic barriers as happen in other oxides [32]. This hypothesis is consistent with the fact that the HP phase predicted by DFT calculations has been observed in CdV$_2$O$_6$ at 6 GPa but at a temperature of 1473 K [33] and in MgV$_2$O$_6$ at 5 GPa and 1273 K [34].

Regarding, the different evolution with pressure of different Bragg peaks, we found that it is caused by an anisotropic behavior of ZnV$_2$O$_6$. As we will show next, the axial compressibilities of *a*, *b*, and *c* are very different. From the structural refinements, we determined the pressure dependence of unit-cell parameters. The results are shown in Fig. 6. There it can be seen that the *a*- and *c*-axis are much more compressible than the *b*-axis. In addition, we found that the β angle follows a non-linear dependence with pressure. Being the crystal structure of $ZnV_2O_6$ monoclinic, its compressibility tensor is a second rank symmetric tensor with only four elements different to zero [35]. Therefore, to discuss the changes induced by pressure in the crystal structure, it is needed to diagonalize this tensor, determining the magnitudes and the directions of the principal axes of the compressibility tensor [36,37]. The results obtained using PASCal [38] are shown in Table 1. In this table it can be seen that the *b*-axis corresponded to the direction of minimum compressibility, being the compressibility one order of magnitude to the direction of maximum compressibility, which is perpendicular to the *b*-axis and making a 60º angle going from the *a*-axis to the *c*-axis. The contraction of the crystal structure along this direction is favor by the layered structure of $ZnV_2O_6$ (see Fig. 1) described in the previous section. The low compressibility along the *b*-axis is a consequence of the linear chains of $VO_4$ and $ZnO_6$ polyhedra aligned along this direction.

In Fig. 7 we report the pressure dependence of the unit-cell volume. The results are compared with previous experiments [11] and with density-functional theory calculations [10]. The first conclusion is that our volume is systematically larger than the volume reported by Tang et al. [11]. However, our volume agrees better with the known ambient-pressure volume that their volume; which underestimates it by more than 1%. In addition, their results show a larger data scattering. When comparing with calculations, we found that calculations using the B3LYP potentials [10] and those that reproduce better the experimental results.

We fitted the pressure−volume (P−V) results from our experiments using a third-order Birch−Murnaghan (BM) equation of state (EOS) [39]. The values of the zero-pressure volume ($V_0$), bulk modulus ($K_0$), and its pressure ($K_0'$) are summarized in Table 2. The value obtained for $K_0$' (which is consistent with 4 within error bars) indicate that our results can be well described by a second-order BM EOS. Comparing with previous results (see Table 2), it can be seen that our bulk modulus is 15% smaller the one determined from calculations. Such differences are within the typical differences between DFT calculations and experiments [40,41]. In the previous experimental study [11] a bulk modulus similar to the calculated values was reported (see Table 2). However, the second-order BM EOS reported in Ref. [11] poorly describes the experimental results as can be seen in Fig. 7 (compare red solid line with black solid squares). In particular, the EOS from Ref. [11] underestimates the ambient-pressure volume and the volume of the first pressure point, which are the most important volumes for determining $K_0$. In addition, the same EOS [11] does not accurately reproduce the dependence of the experimental results with a maximum pressure difference between experiments and the EOS of 1.5 GPa (see Fig. 7). We have therefore, fit the results from Ref. [11] using a third-order BM EOS. The obtained fit is shown with a purple solid line in Fig. 7. In this figure it can be seen that the third-order EOS reproduces much accurately the experiments from Ref. [11]. The obtained EOS parameters are given in Table 2. A bulk modulus 10% smaller than in our experiments is obtained. This is consistent with the faster decrease of the volume with pressure in the results from Ref. [11] and confirms the tendency of calculations to overestimate $K_0$. The small differences between the bulk modulus obtained from the two experiments – 129(2) GPa and 117(5) GPa - can be related to the use of different pressure media [42] (Ar in one case and 16:3:1 methanol-ethanol-water in the other). Notice, that both values compare well with the bulk obtained from the Zn-O bond distance ($K_0$ = 124

GPa) assuming that the crystal compressibility in a first approximation is dominated by the compressibility of the $ZnO_6$ octahedron, which is much more compressible than the $VO_4$ tetrahedron [43].

To close this section, we will compare the results with $MgV_2O_6$ [44]. For this compound, it was proposed the existence of an isostructural phase transition, evidenced by a compressibility change and by a non-linearity in the pressure dependence of phonons [44]. A bulk modulus of 53 GPa was proposed to describe the results for P < 4 GPa and a bulk modulus of 188 GPa for results from 4 to 32 GPa. Such a change in the bulk modulus is impossible for a second-order isostructural transition. In fact, changes in the compressibility has been observed in other oxides, being caused by subtle structural modifications, but not phase transitions [45]. In fact, we have found that the volume reported for $MgV_2O_6$ follows a very similar pressure dependence than in $ZnV_2O_6$. The pressure dependence of the volume for both compounds is shown in the inset of Fig. 7. The results from $MgV_2O_6$ can be described by a third order BM EOS (see Fig. 7) as the results from $ZnV_2O_6$. We obtained for $MgV_2O_6$, $V_0$ = 200(2) Å$^3$, $K_0$ = 120(9) GPa, and $K_0'$ = 6(1). Notice that this bulk modulus is similar to that of $ZnV_2O_6$ and compare well with that obtained assuming that compressibility is dominated by $ZnO_6$ octahedra ($K_0$ = 124 GPa). This suggests that other isostructural metavanadates are expected to have also a similar bulk modulus.

**$Zn_3V_2O_8$**

We have performed two experiments in this compound up to a maximum pressure of 14.9 GPa. One of them was carried out using Ne as pressure medium and the other using methanol-ethanol water. In Figs. 8 and 9 we show XRD patterns from both experiments. They lead to the same conclusion. At the lowest pressure the XRD pattern can be explained using the known crystal structure of $Zn_3V_2O_8$ [8]. In spite of the presence of preferred orientations, good Rietveld refinements can be obtained at all pressures assuming the atomic positions reported in the literature [8]. The results of the fits are included in the figures. At the highest pressure of each experiment, the XRD pattern can be also explained assuming the same structure as shown by the Rietveld refinements shown in the figure. This means that as in $ZnV_2O_6$, in $Zn_3V_2O_8$ there is no phase transition up to 15 GPa.

From the analysis of the XRD patterns we obtained the pressure dependence of the unit-cell parameters and volume. These results are shown in Figs. 10 and 11. In Fig. 10 it can be seen that the compression is not isotropic. The compresibilities of the different axes are $k_a = 2.9(1)\ 10^{-3}$ GPa, $k_b = 1.9(1)\ 10^{-3}$ GPa, and $k_c = 2.7(1)\ 10^{-3}$ GPa. Regarding the pressure dependence of the volume, we have analyzed it using a $2^{nd}$ and a $3^{rd}$ BM EOS. The results from the fits are shown in Fig. 11 and the obtained parameter are summarized in Table 2. Our results show that $Zn_3V_2O_8$ is more compressible than $ZnV_2O_6$ (the bulk modulus is 7% smaller). This is because there are more $ZnO_6$ octahedra in the unit-cell of the first compound than in the second compound. The $ZnO_6$ octahedron is more compressible than the $VO_4$ tetrahedron. Therefore, it is reasonable that an increase of the content of Zn in the chemical formula with lead to a decrease of the bulk modulus.

## $Zn_2V_2O_7$

We will discuss now the results from the pyrovanadate, which is very different than in the other compounds. In Fig. 12 we show XRD patterns up to 1.08 GPa. Those measured at ambient pressure and at 0.2 GPa can be assigned to the low-pressure phase with space group C2/c. At 0.36 GPa this phase dominates the XRD pattern, but additional peaks are present. At 0.66 GPa the new peaks (some of them are denoted by arrows in the figure) become dominant and those of the ambient pressure phase are weak. This indicates the occurrence of a phase transition. The high-pressure phase can be identified with a monoclinic structure described by space group C2/m which is isostructural to the β-polymorph of $Zn_2V_2O_7$. This structure is shown in Fig. 13. It is formed by $ZnO_6$ octahedra and $VO_4$. It is reasonable to expect that under compression, the α-polymorph with five-coordinated Zn atoms would transform to a polymorph with six-coordinated Zn atoms. In Fig. 12, we show a LeBail fit to the XRD pattern measured at 1.08 GPa showing that the HP structure hear proposed is a reasonable structure for the HP phase of $Zn_2V_2O_7$. In addition to the peaks of this compound, we identified two peaks form the gasket (indicated by G) and a few weak peaks identified with $V_2O_5$ (identified by asterisks). This suggests a partial decomposition of $Zn_2V_2O_7$ as previously in other vanadates [46]. From the intensity of the $V_2O_5$ peaks we estimate a 1% of vanadium oxide present at 1.08 GPa. The unit-cell parameters of the HP phase of $Zn_2V_2O_7$ at 1.08 GPa are: $a = 6.652(5)$ Å, $b = 8.458(6)$ Å, $c = 4.966(5)$ Å, and $β = 106.1(24)°$.

The high-pressure phase is identified up to 2.2 GPa. Additional changes occur in the XRD pattern at 2.97 GPa, indicating a second phase transition (see Fig. 14). The second high-pressure phase is identified with a triclinic structure (space group P-1) isomorphic to $Mg_2V_2O_7$ [47]. $a = 13.62(1)$ Å, $b = 5.234(6)$ Å, $c = 4.923(5)$ Å, and $α = 81.0(2)°$, $β = 107.3(3)$, and $γ = 130.0(4)$.

In Fig. 15 we show XRD pattern measured up to 10.75 GPa. The XRD pattern gradually change, but all of them can be assigned to the triclinic HP phase up to 9.2 GPa. These changes suggest a gradual enhancement of the symmetry of the crystal, with several peaks merging as pressure increase. At 10.75 GPa qualitative changes occur in the XRD pattern indicating the existence of a third phase transition. The pattern measured at 12.25 GPa corresponds to the same phase (see Fig. 15) and the same can be said for all the patterns we measured up to 14.7 GPa. The XRD pattern suggest a crystal structure with a higher symmetry than triclinic, however, we could not identify it yet. The deterioration of the XRD pattern due to the stresses induced by the successive phase transitions preclude the proper identification of the crystal structure of the phase found at 10.75 GPa. The identification of it may request the performance of single-crystal XRD studies [48] which will be goal of a future study.

We will comment know on the compressibility of the volume of the ambient-pressure and first high-pressure polymorphs of $Zn_2V_2O_7$. The results are shown in the inset of Fig. 11. The ambient-pressure phase is extremely compressible, with the volume being reduced 1% in only 0.5 GPa. We have only four pressure point for this phase. Therefore, to minimize the number of fitting parameters, we have used a second-order BM EOS to describe the results. The obtained bulk modulus 58(9) GPa is less than half the bulk modulus of the other studied compounds. The calculated EOS is compare with the experiments in the figure. The large compressibility of $Zn_2V_2O_7$ is related to the connectivity between $ZnO_5$ and $VO_4$ polyhedra, which favor volume contraction not only by reducing the volume of the polyhedra but also by means of polyhedral tilting. In addition, the presence of $ZnO_5$ polyhedra also make easier the volume contraction because compression can be accommodated by reducing the empty space in the crystal by favoring the approximation of the closest oxygen atom to the $ZnO_5$ units, which at ambient pressure is at 3.35 Å from Zn [49].

In many aspects, the behavior of $Zn_2V_2O_7$ under pressure resembles that of other pyrovanadates like $ZrV_2O_7$ [50] which has a similar polyhedral connectivity than $Zn_2V_2O_7$. $ZrV_2O_7$ is also very compressible and undergoes a phase transition at very low-pressure (1.5 GPa) [50]. This suggest that $Zr_4V_2O_9$, with a structural framework more similar $Zn_2V_2O_7$ than to the other vanadates, would be probably very compressible too and undergo phase transitions at low pressure.

In the inset of Fig. 11 we also show the pressure dependence of the volume for the monoclinic HP phase of $Zn_2V_2O_7$. The first remarkable fact is that at the transition there is a volume collapse of 7%. The HP phase is much less compressible than the low-pressure phase as can be seen in the figure. For the HP phase, a second-order EOS leads to a bulk modulus of 151(4) GPa, which is of the same order than the bulk modulus of $ZnV_2O_6$ and $Zr_3V_2O_8$ being consistent with the fact that Zn is six-coordinated by oxygen atoms in the HP phase. For the triclinic HP phase, we have not studied the pressure dependence of the volume, but have noticed that the second transition involves a larger volume collapse than the first transition (~20%).

**Summary**

We report a high-pressure powder x-ray diffraction study of $ZnV_2O_6$, $Zn_2V_2O_7$, and $Zn_3V_2O_8$ up to 15 GPa. We present evidence of the existence of three phase transitions in $Zn_2V_2O_7$. In contrast, no phase transitions have been found in the other compounds. The difference is the high-pressure behavior is discussed in terms of the polyhedral networks of the different vanadates. In addition, we found that $Zn_2V_2O_7$ is much more compressible than the other vanadates, resembling the behavior of other pyrovanadates. Equations of state have been determined for the different compounds. The results are compared with those for related compounds. General conclusions for the high-pressure behavior of zinc vanadates are extracted from this study.

**Acknowledgments**

This work was supported by the Spanish Ministry of Science, Innovation and Universities under Grant MAT2016-75586-C4-1-P and by Generalitat Valenciana under Grant Prometeo/2018/123 (EFIMAT). The authors thank ALBA synchrotron for providing beam-time for the XRD experiments (Projects 2016081779 and 2019023362).

**Table 1**

| | |
|---|---|
| $\lambda_1 = 4.59(5)\ 10^{-3}$ GPa$^{-1}$ | $e_{v1} = (-0.5019, 0, -0.8649)$ |
| $\lambda_2 = 1.29(4)\ 10^{-3}$ GPa$^{-1}$ | $e_{v2} = (0.7424, 0, -0.6699)$ |
| $\lambda_3 = 0.39(1)\ 10^{-3}$ GPa$^{-1}$ | $e_{v3} = (0, 1, 0)$ |

**Table 2**

| ZnV$_2$O$_6$ | $V_0(\text{Å}^3)$ | $K_0(GPa)$ | $K_0'$ |
|---|---|---|---|
| This work | 200.18(6) | 129(2) | 4.1(3) |
| Tang et al. | 197.47 | 147 | 4 (fixed) |
| Refit to Tang el al. | 199(1) | 117(5) | 6.5(5) |
| Theory (B3LYP) | 199.908 | 146.74 | 5.11 |
| Theory (HSE06) | 186.602 | 152.18 | 5.54 |
| Theory (PBE) | 197.219 | 147.39 | 5.09 |
| Zn$_2$V$_2$O$_7$ | $V_0(\text{Å}^3)$ | $K_0(GPa)$ | $K_0'$ |
| This work (LP) | 581.4(6) | 58(9) | 4 (fixed) |
| This work (HP) | 269.6(6) | 151(4) | 4 (fixed) |
| Zn$_2$V$_2$O$_7$ | $V_0(\text{Å}^3)$ | $K_0(GPa)$ | $K_0'$ |
| This work | 585.1(1) | 120(2) | 4 (fixed) |
| This work | 585.0(4) | 115(2) | 5.1(6) |

**Figure 1**

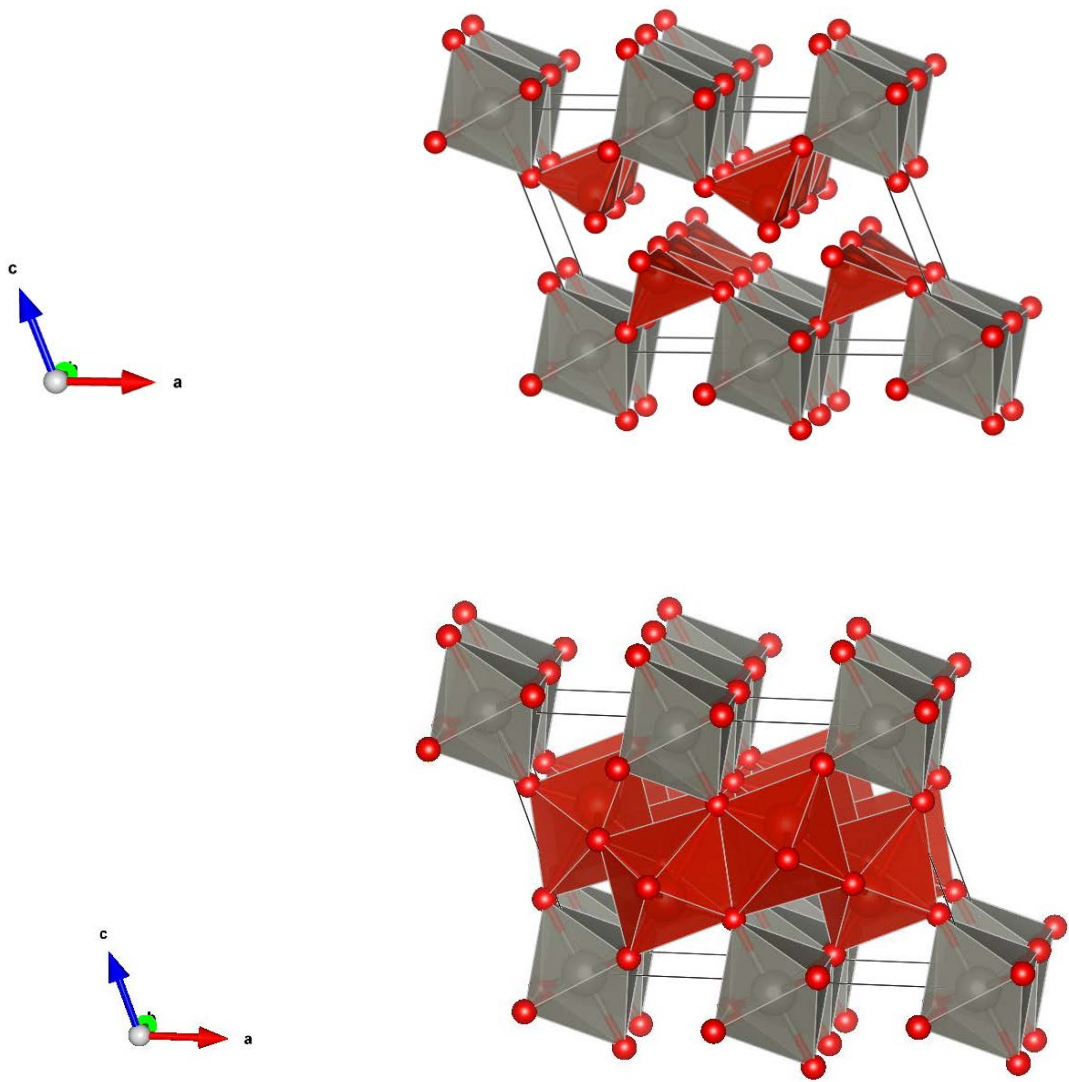

**Figure 2**

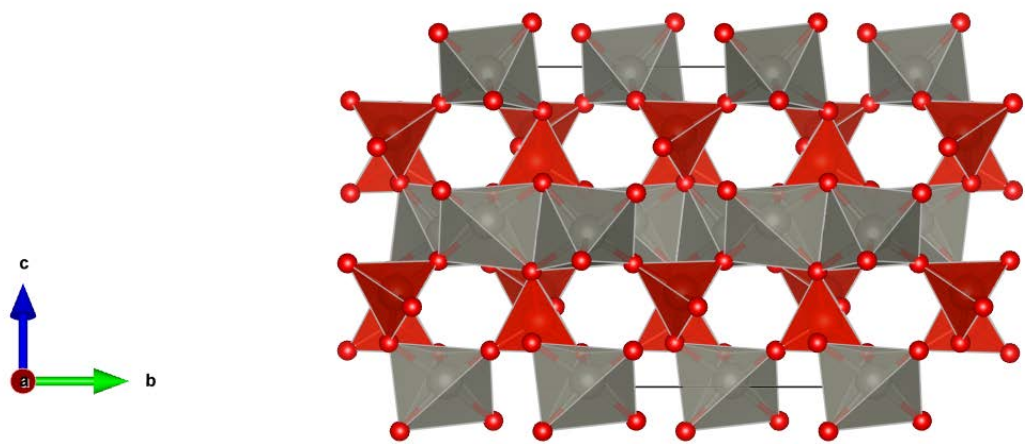

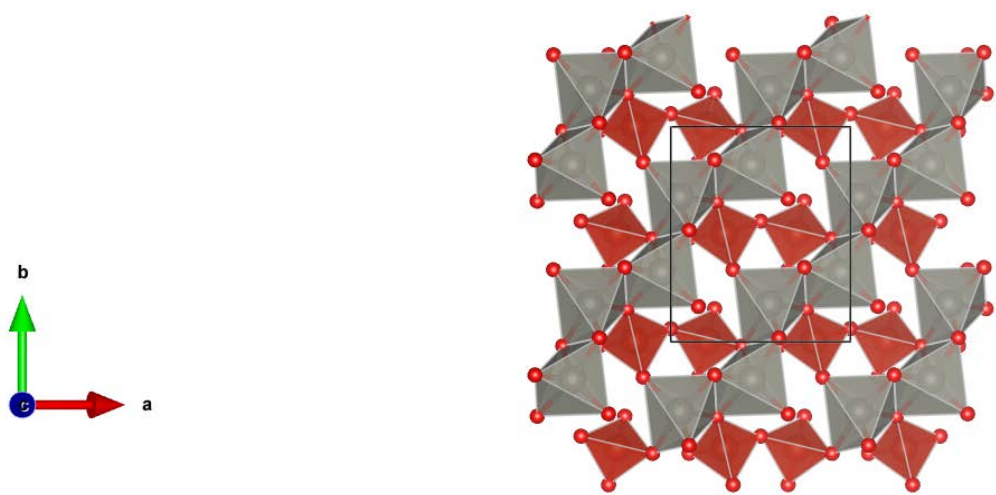

**Figure 3**

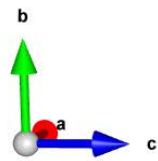 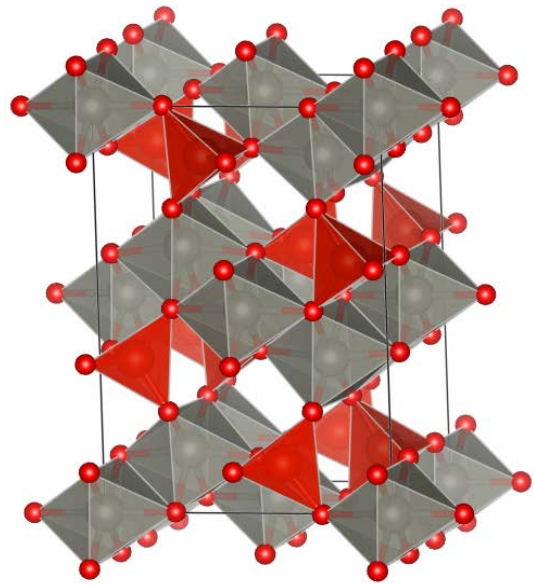

**Figure 4**

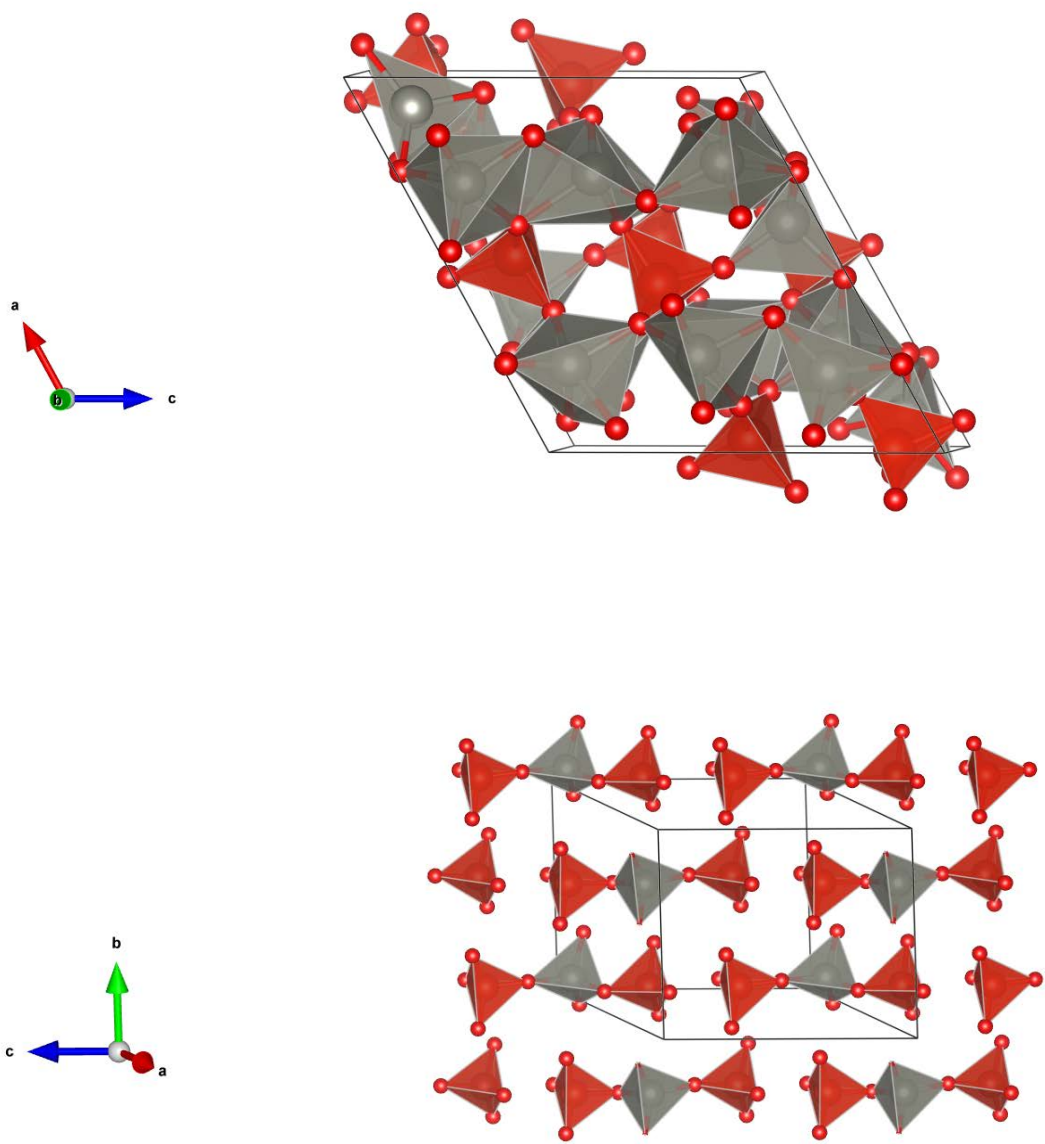

**Figure 5**

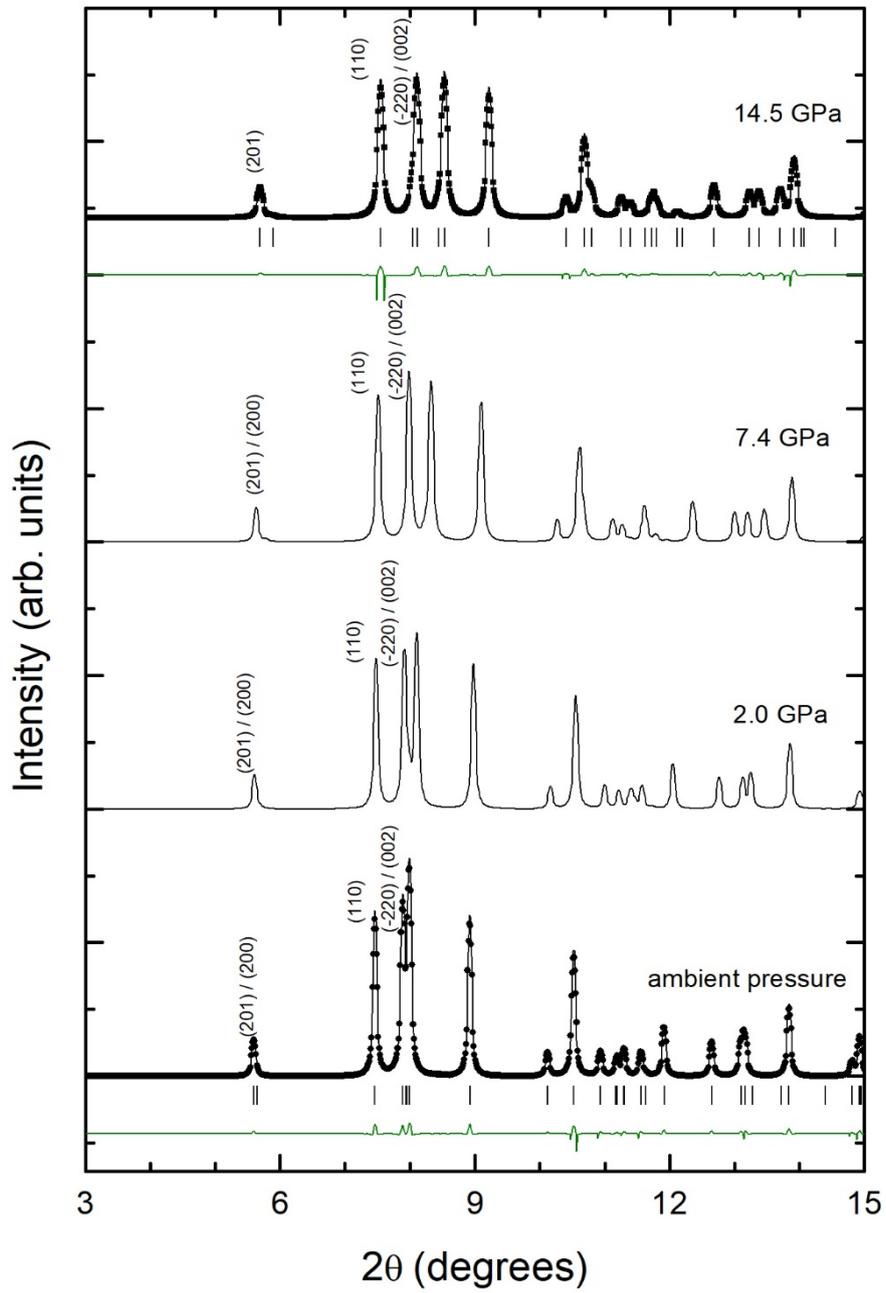

**Figure 6**

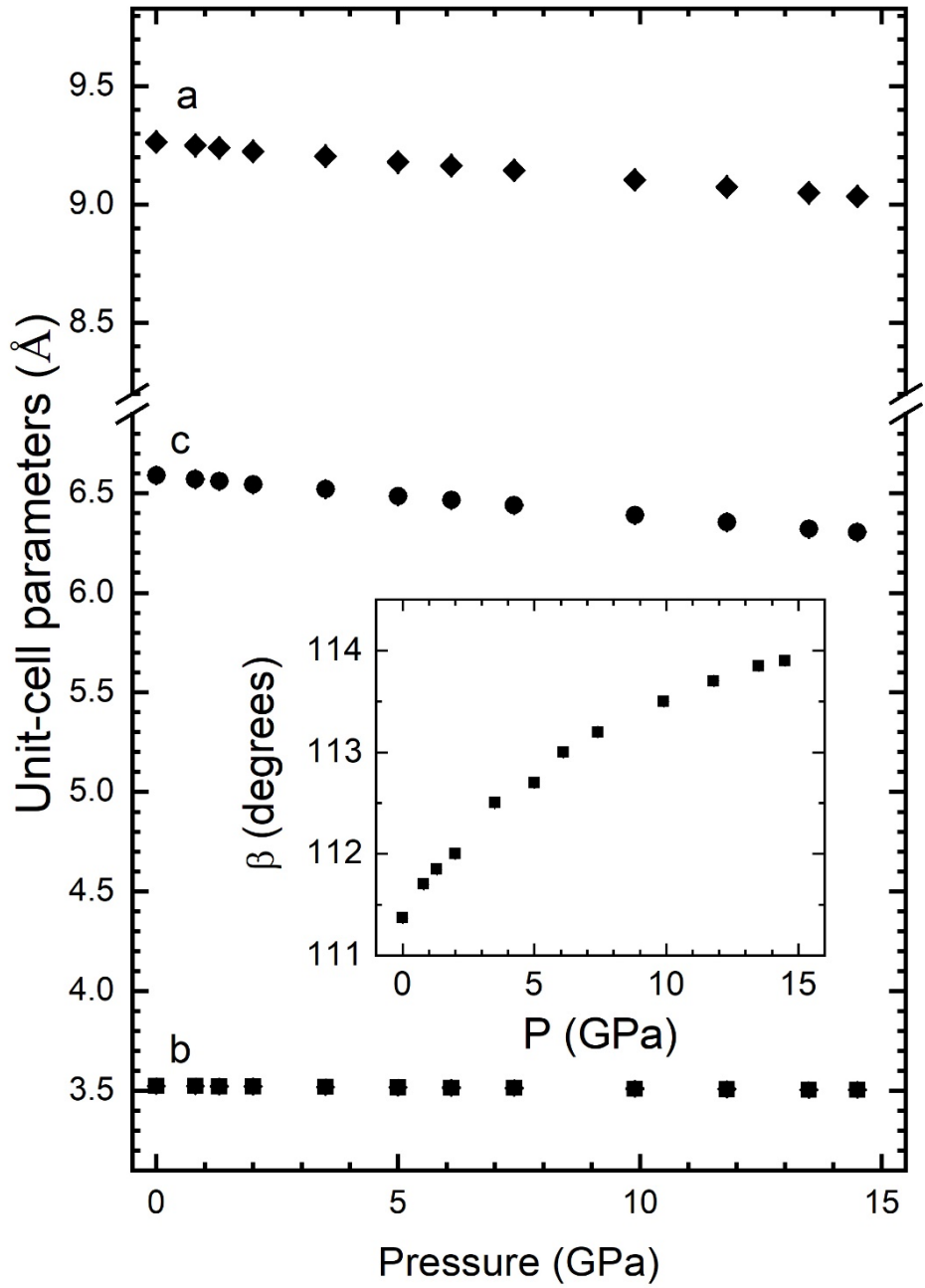

**Figure 7**

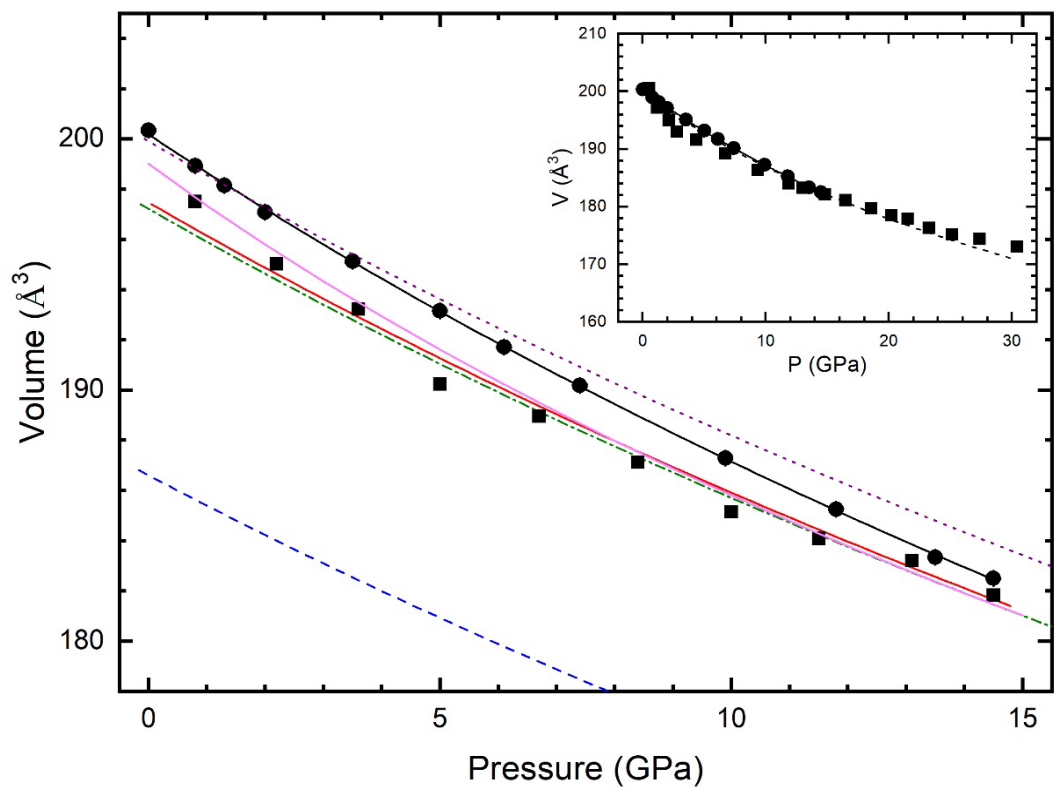

**Figure 8**

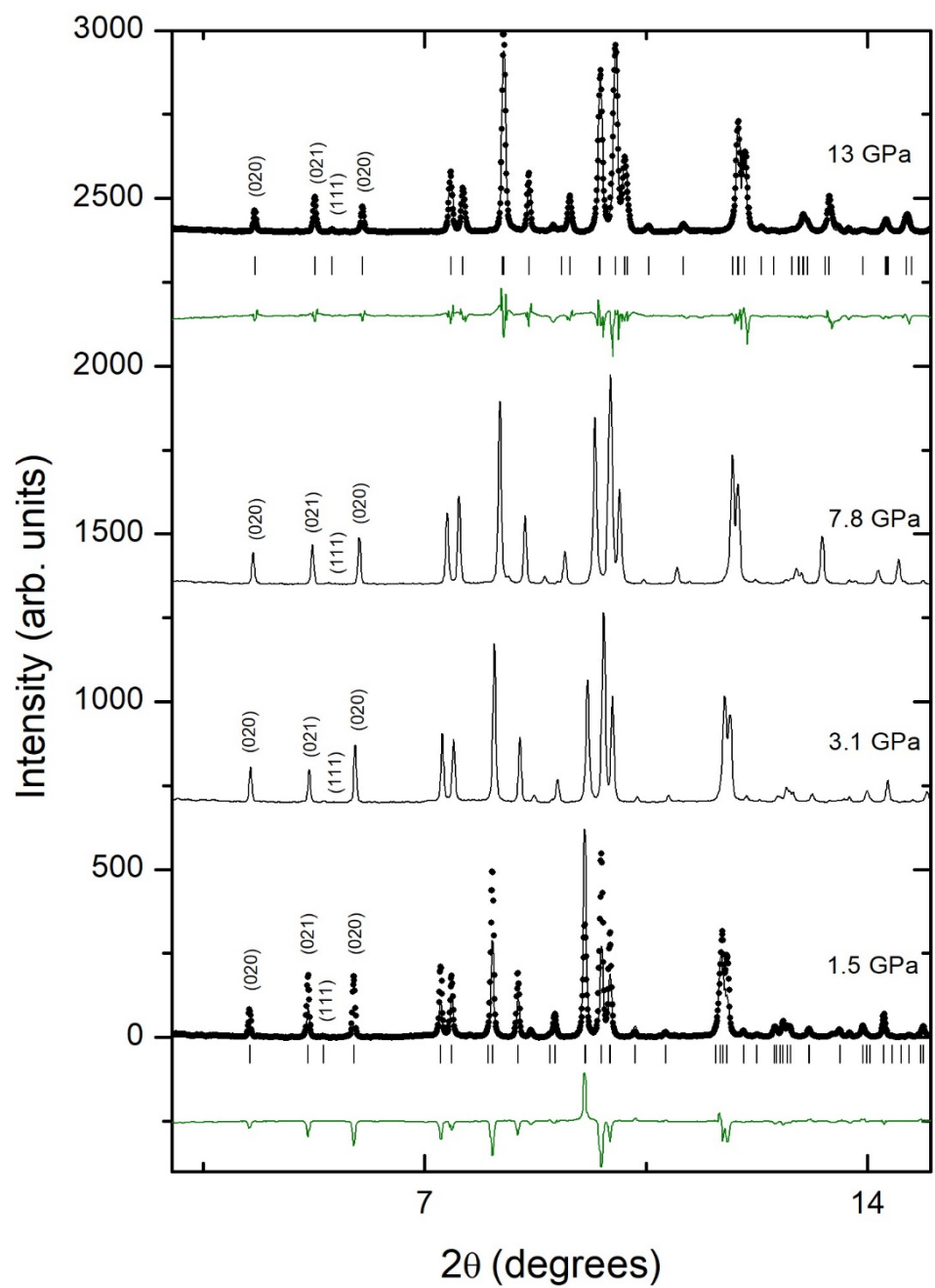

**Figure 9**

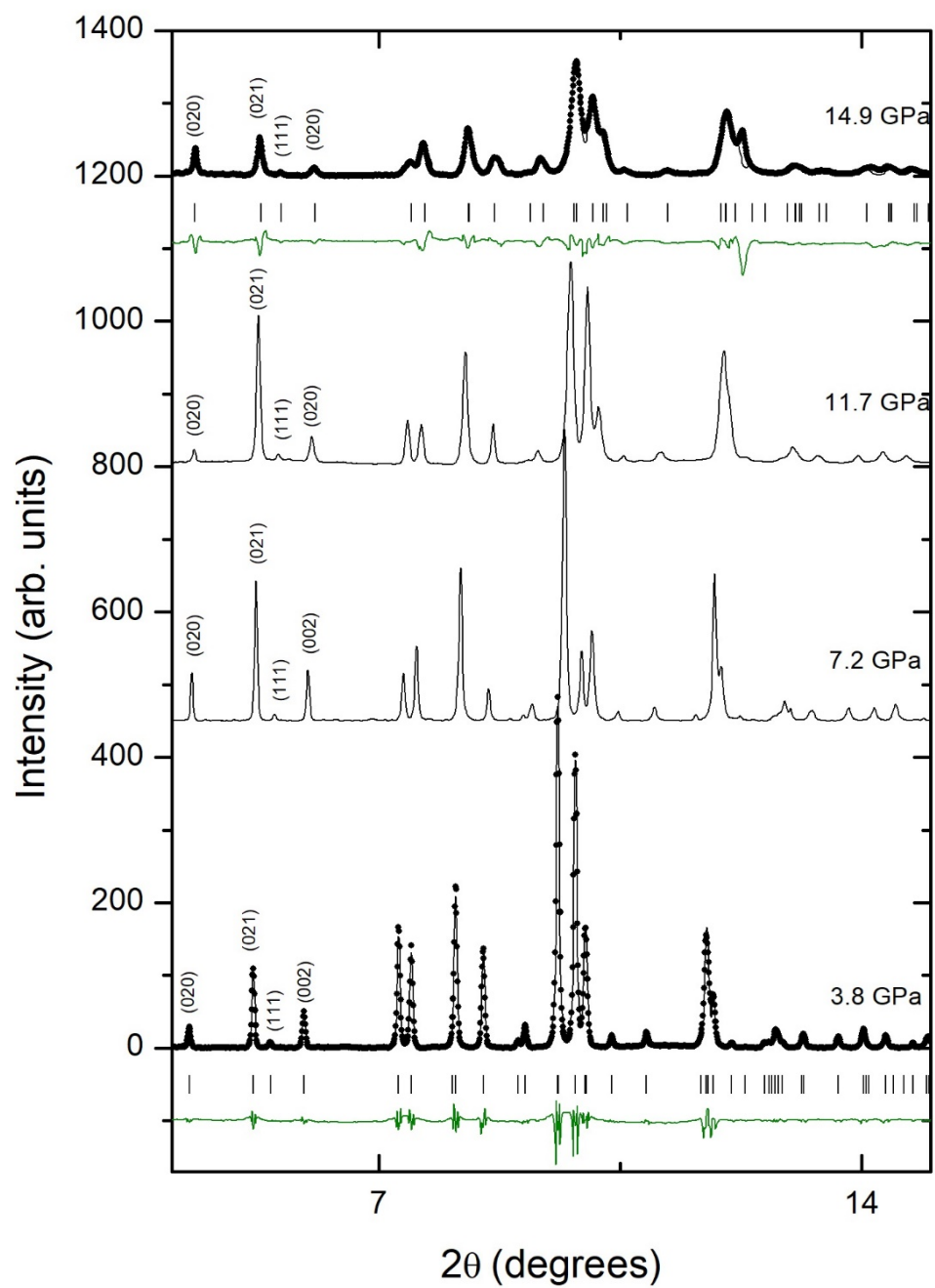

**Figure 10**

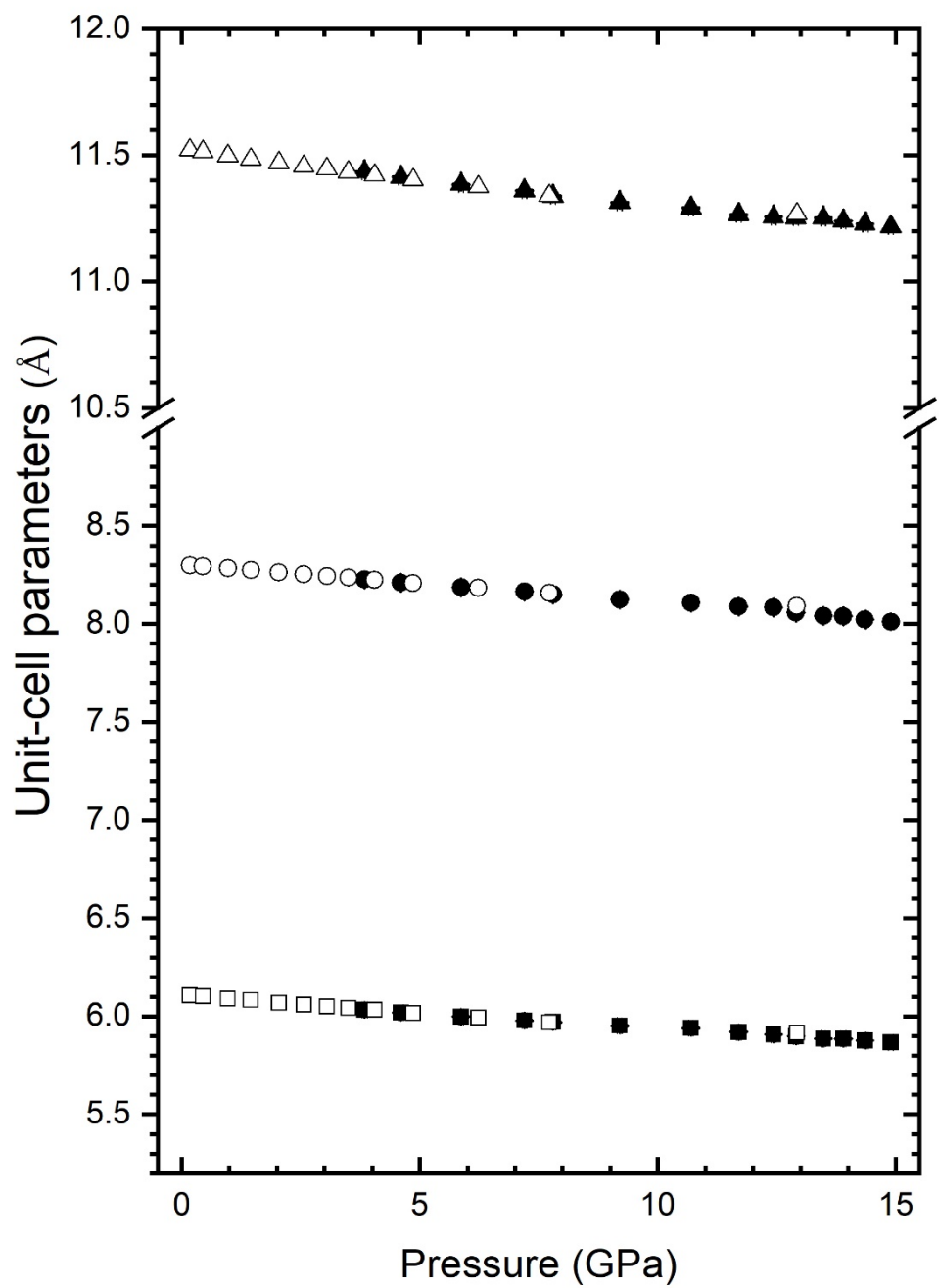

**Figure 11**

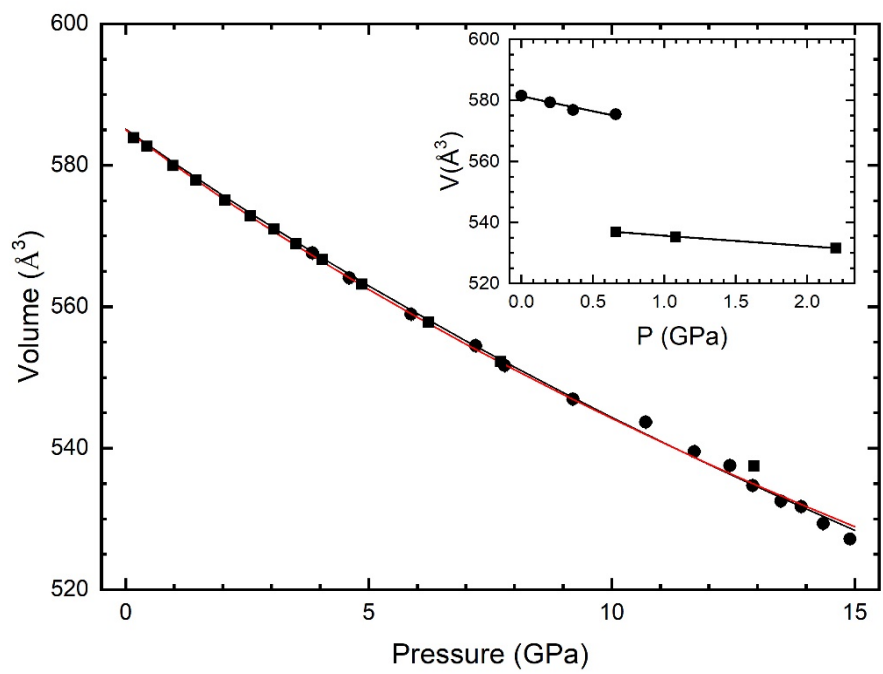

**Figure 12**

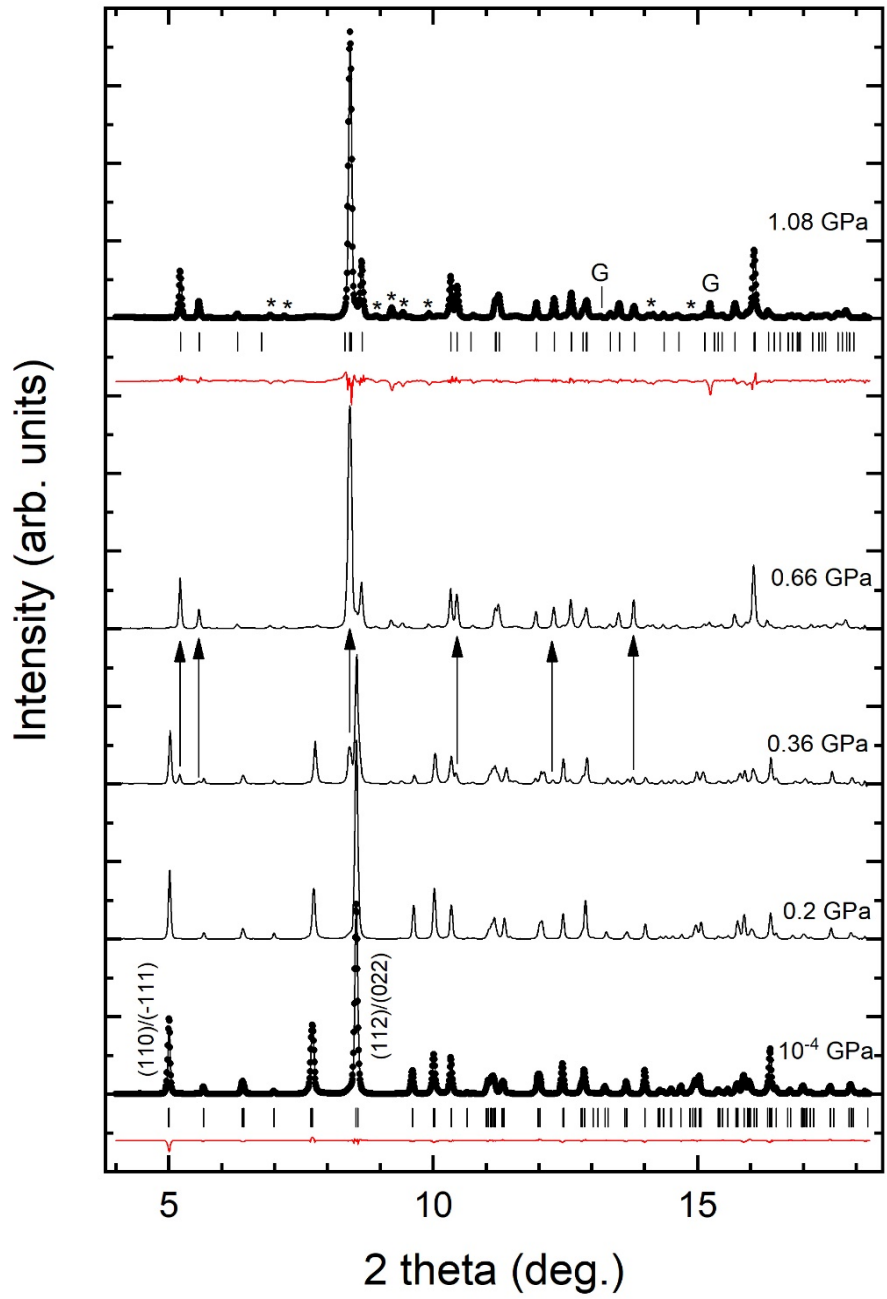

**Figure 13**

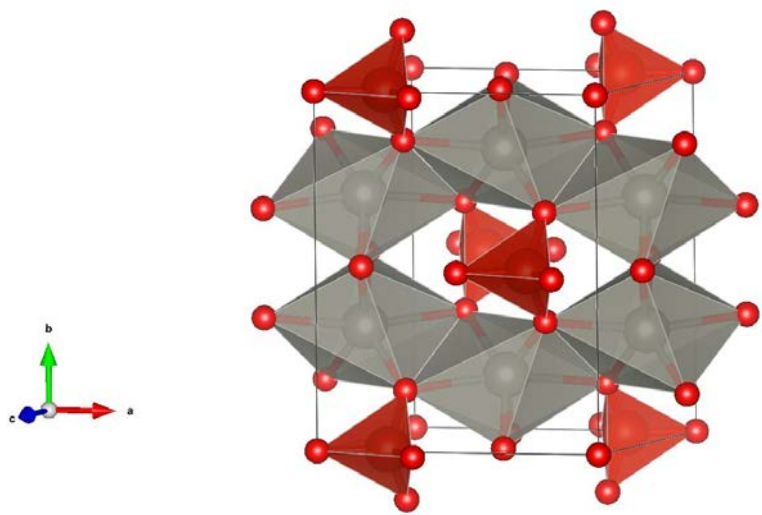

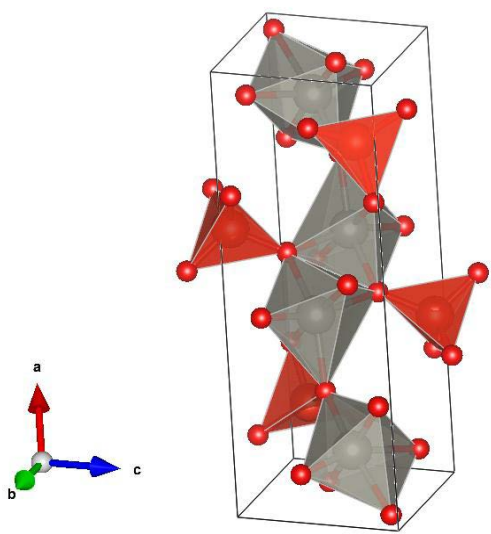

**Figure 14**

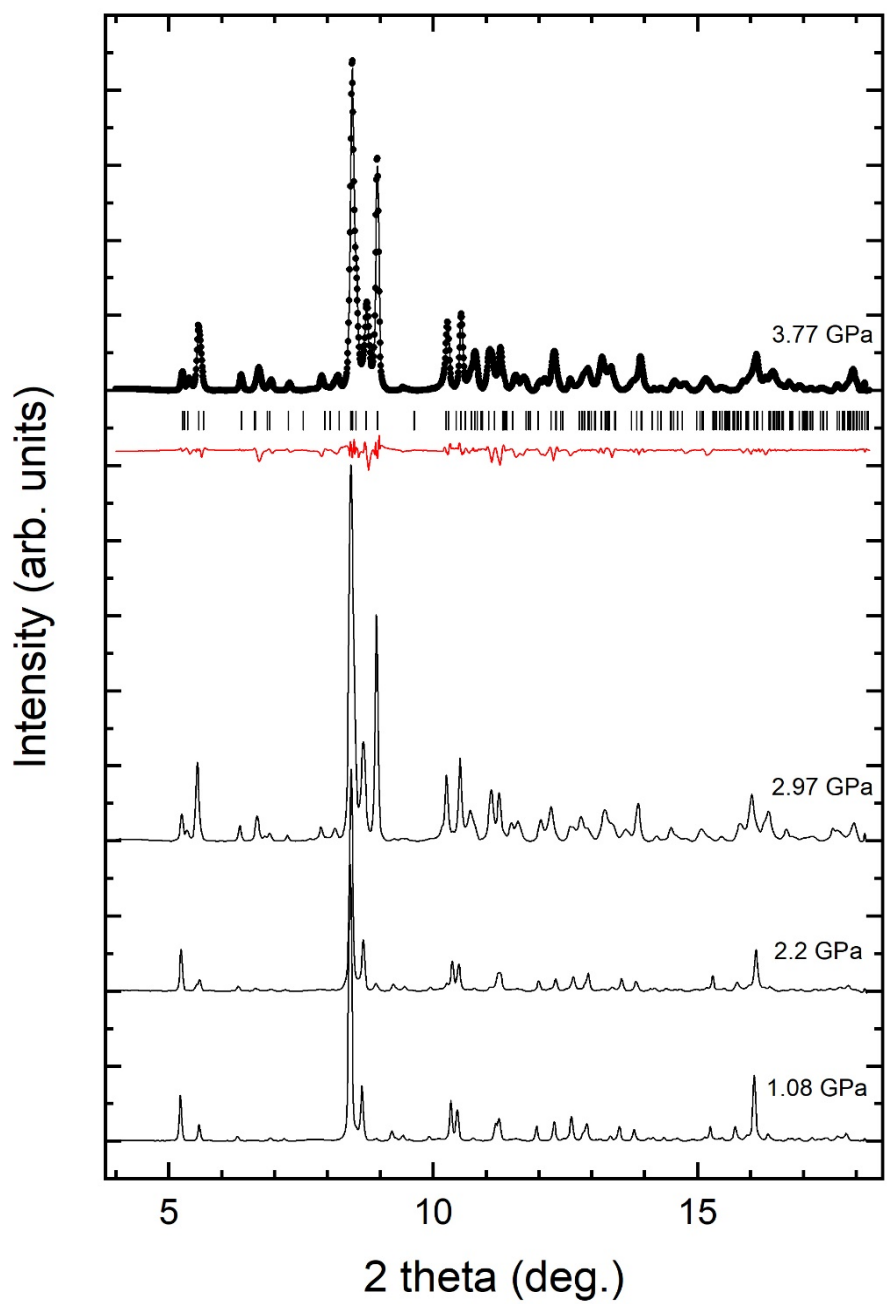

**Figure 15**

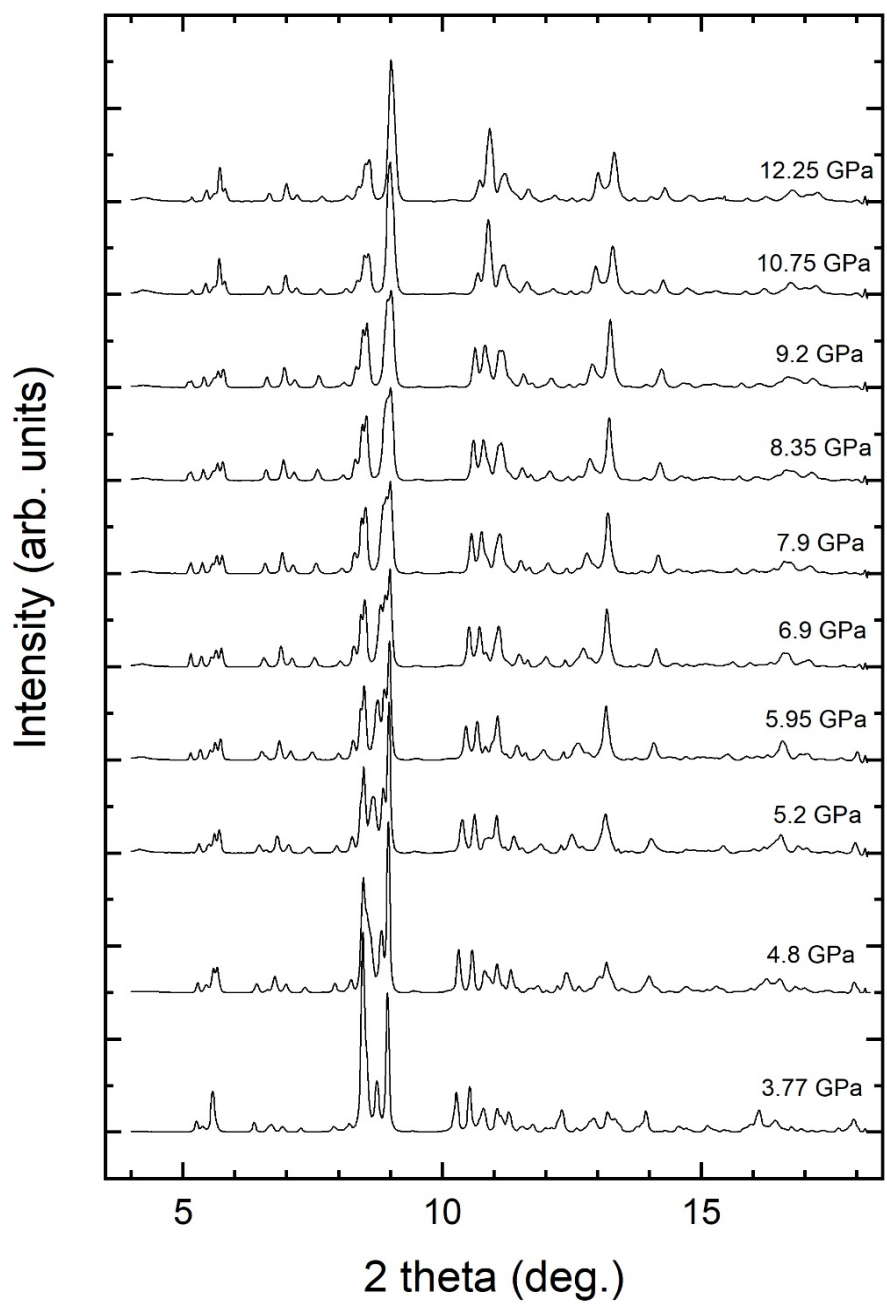